# Fire acting as an increasing spatial autocorrelation force: Implications for pattern formation and ecological facilitation

Aristides Moustakas[1, *]

1. School of Biological and Chemical Sciences

   Queen Mary University of London

   Mile End Road, E1 4NS, London, UK

   * Corresponding author

   Email: arismoustakas@gmail.com

**Abstract**

Fire is an indissoluble component of ecosystems, however quantifying the effects of fire on vegetation is challenging task as fire lies outside the typical experimental design attributes. A recent simulation study showed that under increased fire regimes positive tree-tree interactions were recorded . Data from experimental burning plots in an African savanna, the Kruger National Park, were collected across unburnt and annual burn plots. Indices of aggregation and spatial autocorrelation of the distribution of trees between different fire regimes were explored. Results show that the distribution of trees under fire were more clumped and exhibited higher spatial autocorrelation than in unburnt plots: In burnt plots spatial autocorrelation values were positive at finer scales and negative at coarser scales potentially indicating co-existence of facilitation and competition within the same ecosystem depending on the scale. The pattern derived here provides inference for (a) fire acting as an increasing aggregation & spatial autocorrelation force (b) tree survival under fire regimes is potentially facilitated by forming patches of trees (c) scale-dependent facilitation and competition coexisting within the same ecosystem with finer scale facilitation and coarser scale competition.

# Introduction

Fire is and has historically been an important regulator of tree community structure in different ecosystems . In most cases fires are hard to predict and to control, and their spatial scale exhibits high variance spanning to different levels of magnitude of hectares. Thus, fires lie outside the typical examples of experimental design, and as a result analyzing their effects is often a hard task . Theoretical studies have indicated that the spread of fire can be critically influenced by the density as well as the spatial structure of trees .

Tree spacing can be random, regular, or aggregated (also known as clumped). The regular pattern is created by competition between aggregated neighbouring individuals and death of some of them . In the absence of fire, clumped distributions can be formed by anthropogenic disturbances, soil patchiness, vegetative reproduction, limited dispersal capabilities, as well as gap regeneration - see and references therein. Theoretically, assuming spatial homogeneity, the spatial distribution that provides optimal growth opportunities for all trees is achieved by equal tree spacing, i.e. regular tree distribution . It has been reported that spatial autocorrelation did not greatly influence assessments of fire effects and that treatments designed to assess heterogeneity in forest conditions prior to the reintroduction of fire will likely be unnecessary . However, a recent study reports increased clumping of trees under more frequent burning, and thus providing inference for higher spatial autocorrelation in the spatial distribution of trees under more frequent burning .

Often unaccounted for in the past, positive (facilitative) and negative (competitive) plant interactions are now considered to have serious implications for population dynamics and ecosystem function . Ecological facilitation and reciprocally competition has been often investigated in terms of the stress gradient hypothesis , which predicts an increase of positive interactions with increasing environmental stress . To date there are very few studies investigating the effects of fire frequency as a stress gradient on competitive-facilitative interactions on plants and the effects of fire are not necessarily straightforward:  Competition with established vegetation limits plant recruitment and fires create open conditions which can allow new species to establish . However, facilitation theory suggests that destruction of vegetation will restrict germination and survival . A theoretical study recently reported that fire frequency can increase positive tree-tree interactions .

Savannas are ecosystems comprising of a mixture of woody species (trees and bushes), grasses and forbs. They cover about a fifth of the global land surface and about half of the area of Africa, Australia and South America . Savannas are characterized by a continuous grass and a discontinuous tree layer. A savanna, where trees and grasses coexist , may be viewed as an intermediate ecosystem between grassland (grass dominance) and forest (tree dominance) with increasing precipitation resulting in a denser tree layer yet still discontinues in comparison to a forest . Fire is an indissoluble component of savanna ecosystems . Savannas are mainly characterised by ground fires as fuel load derives mainly from grass biomass and such fires occur mainly in mesic or humid savannas .

The aim of this work is to quantify the effects of fire on tree size class distributions and spacing patterns and link pattern and process regarding tree-tree interactions . To that end size and spatial properties of trees in prescribed burning plots in a mesic savanna were compared with unburnt plots on the same location . Prescribed fires managed for resource objectives displayed similar patterns of fire severity, heterogeneity, and seedling and sapling survival with wildfires, creating post-fire conditions that approximate natural fires when assessed on a fine spatial scale . Using data from experimental burning plots that included annual burning vs. no burning (control) for the past ~50 years .

The properties of trees examined included indices of aggregation as well as the spatial autocorrelation of their distributions. Two questions were addressed in this study: (i) Are trees more aggregated in burnt than in unburnt treatments? This could be the result of fire-generated open conditions resulting in lower tree occupancy thereby leading to a higher aggregation  Alternatively a higher aggregation in burnt plots could result from fire increasing positive tree-tree interactions  due to an increased stress gradient  with fire as the stressor. (ii) Are tree distributions exhibiting higher spatial autocorrelation values in burnt plots? This would follow up from a potential higher aggregation in burnt plots .

**Methods**

Data were collected in the Kruger National Park (KNP), South Africa between February and March 2010. The park is situated in the savannas of north-eastern

South Africa (Fig. 1), and covers an area of ~19,633 km$^2$. The vegetation in the park is mainly characterized by dense savanna dominated by Acacia and Combretum species. Within the park there are long-term Experimental Burning Plots (EBPs) where fire is manipulated as a treatment for the past ~50 y and thus KNP is an ideal environment for comparing fire effects on vegetation . The experiment was laid out between 1954 and 1956. In 1979, some plots, in some landscapes were split to allow the implementation of additional treatments ; no data are analyzed from these new treatments. The experiment is replicated four times within the Skukuza landscape (plots) within the park. The Skukuza landscapes are on granite soils and the mean annual precipitation is 550 mm. Each replicate plot consists of two different experimental treatments and each treatment is implemented in a 7-ha plot in a split-plot randomised design ; (Fig 2a). The burning treatment includes annual burning every August (*F1*) which is the dry season, while the control treatment excludes fire (*F0*).

In each of the eight plots, four line transects of 200 m length each were drawn. Each transect was distanced 40 m from the previous/next transect. For each transect, every 20 m (focal points) the nearest two adult individuals to the focal points (focal trees) were recorded (Fig. 2b). Thus 10 trees per focal point, 100 trees per transect and 400 trees per plot were sampled. This resulted in 400 trees per plot x 8 plots = 3200 trees in total were sampled, 1600 trees in unburnt, and 1600 in annual burn plots. There were a few cases where the one of the four nearest neighbours of focal tree1 was also one of the four nearest neighbours of focal tree2 – in those cases the duplicate tree individuals were only included once in the statistical analysis.

Recorded variables per tree sampled included species identity, Trunk Circumference at Breast Height (TCAH, measured at 1.35 m above ground), and the X, Y coordinates of the trunk of each focal tree. Only adult (i.e. reproducing – no saplings or seedlings) individuals were sampled. The species ID, TCAH, and X, Y coordinates of the four nearest neighbours of each of the two focal trees were also sampled. Dominant species identities included mainly *Acacia nigrescens*, *Dichrostachys cinerea*, *Sclerocarya birrea*, *Combretum collinum*, *Combretum apiculatum*, *Terminalia sericea* and *Euclea divonorum*. All necessary permits for field work were obtained from the Administration, Scientific Services and the local Rangers of the Kruger National Park.

*Analysis*

For each tree recorded within each plot and burning treatment (*F0*, *F1*), the average of the distances to the four nearest neighbouring trees was recorded, based on the inter-trunk distances of each pair of trees for each burning treatment (*MeandistF0*, *MeandistF1*). The distance to the nearest neighbouring tree for each burning treatment was also recorded (*NearestF0*, *NearestF1*). The mean size in terms of TCAH of the nearest four tree individuals was recorded (*MeansizeF0*, *MeansizeF1*). The TCAH of the largest among the four nearest tree individuals was also recorded (*LargestF0*, *LargestF1*). The coefficient of variation (CV), skewness, and kurtosis of the summary statistics of the abovementioned size and distance indices were calculated. The CV is a normalised measure of dispersion of the data expressed as the ratio of the standard deviation to the mean of the data commonly used to assess data aggregation . Skewness is a unit-less index that quantifies how symmetric a distribution is – an asymmetrical distribution with long tail to the right has a positive skew . Kurtosis is a unit-less index that quantifies whether the shape of the data matches a Gaussian distribution and distributions more peaked than a Gaussian have positive kurtosis values .

In the study of spatial patterns and processes, it is logical to expect that close observations are more likely to be similar than those far apart – spatial autocorrelation . In order to assess the spatial autocorrelation of tree size distributions, in terms of TCAH, across distances the Moran's I was used . Moran's I is commonly employed for quantifying characteristics and autocorrelation in the spatial distribution of data . Moran's $I$ is defined as:

$$I = \frac{N}{\sum_i \sum_j W_{ij}} \frac{\sum_i \sum_j W_{ij} (X_i - \bar{X})(X_j - \bar{X})}{\sum_i (X_i - \bar{X})^2}$$

where N is the number of trees within each plot, X is the TCAH of each tree in the plot, $\bar{X}$ is the mean value of TCAH of trees within each plots, $w_{ij}$ are the weights of each pair-wise ($X_i$, $X_j$) comparison, and i, j are the different locations within each plot that a tree was recorded. The weights of pair-wise comparisons take values closer to 1 for very close neighbours and closer to 0 for very distant neighbours, also known as a neighbouring function . The formula used to calculate the weights was: weights $w_{i,j}$ = $1/d_{i,j}$, where $d_{i,j}$ is the Euclidean distance between points i and j. A permutation test for Moran's I statistic calculated by using 999 random Monte Carlo permutations for the given spatial weighting scheme to establish the rank of the observed statistic in relation to the simulated values was employed . The Monte Carlo-generated pseudo-p values of the test were used to assess significance .The analysis was performed on each of the eight plots and the distant-specific autocorrelation I values for each of the four unburnt and the four annual burn plots were averaged separately and plotted. Analysis was conducted in the 'spdep' package in R .

**Results**

Dominant tree species had similar presences in unburnt and annual burn plots (Fig. 3). The CV, skewness, and kurtosis of the four nearest and in particular the nearest neighbour were larger in annual burning plots indicating a much higher degree of aggregation (clumping) in the spatial distribution of trees (Table 1; Fig. 4). The CV of TCAH of the four nearest as well as the largest of the four nearest neighbouring trees was marginally larger in annual burn than in unburned plots (Table 1), however skewness and kurtosis were larger in unburned than in annual burn plots (Table 1).

Values of Moran's I across distances indicated no spatial autocorrelation of tree distributions within unburned plots at all distances examined up to 120 and repulsion (negative correlation) at long distances above 120 (Fig. 5a). Contrastingly, there was significantly positive spatial autocorrelation in the spatial distribution of trees within annual burning plots for distances up to 30 m, while the spatial autocorrelation of larger distances could not be differentiated from random at distances up to 90 m and negative spatial autocorrelation after 110 m distances (Fig. 5b).

**Discussion**

This study showed that there is increased aggregation in terms of all three indices of aggregation, CV, skewness, and kurtosis, as well as increased spatial autocorrelation (Moran's I) of trees under annual burning treatment. The mean distance to the four nearest neighbours is higher under annual burning than in unburnt plots; However, the higher aggregation and spatial autocorrelation in the distribution of trees implies that trees form patches in the annual burn plots and thus proximity to other neighbouring trees facilitates survival . Fire treatments increased spatial autocorrelation of tree distributions as assessed by Moran's I, an index showing direction (positive or negative) and degree (strength) of spatial autocorrelation . Increased spatial autocorrelation should be interpreted here as deviation from zero or random spatial autocorrelation: Values of Moran's I that are negative and significantly deviating from random are also indicating increased spatial autocorrelation (i.e. not only positive Moran's I values); .

Within the experimental burning plots of the Kruger it has been reported that the density of woody individuals was unresponsive to fire, but the dominance of small trees was highly responsive to the fire regime . In this study only adult individuals were sampled and thus the results reported here in should be interpreted bearing that in mind. In the same study it was reported that savannas are demographically resilient to fire but structurally responsive ; results derived here and  suggest that tree-tree facilitation can be one of the structural responses under fire. The analysis of clustering conducted here did not explicitly account for species identities of neighbouring trees. However species between burnt and unburnt plots did not differ as savanna species are adopted to fire and thus species that are highly flammable and that in theory could survive in burning plots are not present .

Tree aggregation under repeated fire has been reported to initially increase clumping although repeated fires reduced it . The dataset analysed here has no time replicate and although spatial variation can be used as a tool for inferring temporal variation in population dynamics this is not within the scope if this study. However, the experimental burning plots have been established for ~50 y and thus such temporal effects may long have stabilised. Another study comparing active fire and fire suppressed areas indicated that fire exclusion increased patch sizes of trees and decreased spatial complexity , a result in agreement with the findings reported here.

Results derived here show that both positive and negative spatial autocorrelation exist in burnt plots and positive autocorrelation exists at finer scales while negative at larger ones as also reported in an experimental study . It has also been reported that facilitation and competition may coexist within the same ecosystem depending on the scale of analysis with finer scale facilitation and larger scale competition . Combined results derived here in terms of increased aggregation and positive Moran's I at finer scales and negative at larger spatial scales are showing a similar pattern with the one derived by – a fine scale facilitation and a relatively larger scale competition. It would be interesting to elaborate on the scale that this shift potentially occurs and link this scale-specific shift to the underlying mechanisms (stress gradient, soil properties, density, etc) that may be causing it: The shift from facilitation to competition may be occurring at coarser scales as precipitation increases or at finer scales as density increases. In that case quantifying the range and extend of local clustering could explain local facilitation . It is also important to note that theoretical studies indicate that, in the absence of facilitation, non-local competition among plants alone may induce patterns .

This study investigated the effects of fire as a stress gradient across two extremes: no fire for the past 50 y and annual burn during the same period. Most cases that fire acts as a stress gradient are in between those two extremes (very few natural ecosystems get burned every year); for example in the Kruger the mean natural fire return interval is between four to seven years . The frequency of burning is inversely related to the median fuel load (Fig. 6a) but that doesn't necessarily imply a linearly increasing relationship with fire intensity (Fig. 6b). Frequent burning does not build up big wildfires and to that end the effects of big wildfires on the spatial patterning of vegetation across finer and coarser scales may differ from the ones derived here.

How is the pattern derived here formed? In the absence of fire it is well recorded through long-term studies that savanna trees are competing with their nearest tree neighbours: Probability of mortality is increasing with increasing proximity to the nearest neighbours , implying a finer scale competition rather than facilitation. Experiments that manipulated removal of nearest woody species neighbours also reported such competitive effects . In this study crowding increases the chance of survival around tree clumps in annual burning plots. Fire in the study area and in savannas in general is mainly occurring through ground fires, the main fuel load is composed by grass biomass during the dry season (August, when the annual burning treatment occurs). Facilitation can occur through various mechanisms including refuge from physical stress , refuge from predation , refuge from competition , and improved resource availability . While refuge from competition seems unlikely – there is no process why trees should not compete with their nearest neighbours in the presence of fire - the rest are testable feasible explanations: (i) Refuge from physical stress could occur if dense tree patches have lower below canopy grass biomass than in the open and thus chance of fire or fire intensity will be lower. (ii) Refuge from predation could occur because there is higher post-fire grazing intensity though this effect should be less pronounced in adult trees sampled here. (iii) Higher resource availability could occur through higher hydraulic lift .

    Several factors have been proposed to modulate facilitative-competitive interactions in terms of stress gradients: seasonal water availability with competition switching to facilitation within the same year when water resources become scarce , mean annual water availability grazing intensity , and plant density . Clearly there is an interaction effect between fire as a stress gradient and all those factors. Competitive-facilitative interactions across multiple stress gradients are not necessarily linear and may often cancel each other and such studies remain scarce. It would thus be challenging to disentangle the effects of fire in combination with MAP, density, and grazing (multiple stress gradients) on positive or negative interactions in plant communities. In addition this study investigates tree-tree patterns in unburned and annual burned plots. In the same study area (Skukuza in KNP), and in the absence of fire (control plots only) and density (isolated trees sampled only) positive interactions between trees and grasses were reported . It

would be interesting to explore the relationship between tree-grass interactions under fire as a stress gradient.

**Acknowledgements**

I thank Justin Dohn & Niall Hanan for several discussions on the effects of fire on competitive-facilitative interactions, and Mahesh Sankaran & Bill Kunin for suggestions regarding the experimental sampling design. I am grateful to Johan Baloyi, Elliot Thekiso Shilote, Varun Varma, and Fanis Missirlis for invaluable help and company during fieldwork. Comments of two anonymous reviewers considerably improved an earlier manuscript draft. This research was funded by a NERC Research Grant (NE-E017436-1).

| Variable | CoefVar | Skewness | Kurtosis |
|---|---|---|---|
| MeandistF0 | 50.92 | 1.31 | 1.89 |
| MeandistF1 | 67.53 | 2.12 | 6.71 |
| NearestF0 | 56.95 | 0.98 | 0.55 |
| NearestF1 | 84.68 | 2.68 | 12.01 |
| MeansizeF0 | 51.95 | 1.33 | 1.75 |
| MeansizeF1 | 54.25 | 1.07 | 1.06 |
| LargestF0 | 60.35 | 1.99 | 6.71 |
| LargestF1 | 75.65 | 2.08 | 5.43 |

**Table 1.** Values of the coefficient of variation (CV), skewness, and kurtosis of the distances and sizes of neighbouring trees. For each tree recorded within each plot and burning treatment (*F0*, *F1*), the average of the distances to the four nearest neighbouring trees (*MeandistF0*, *MeandistF1*), and the distance to the nearest neighbouring tree (*NearestF0*, *NearestF1*) were recorded. The mean size in terms of TCAH of the nearest four tree individuals (*MeansizeF0*, *MeansizeF1*), and the TCAH of the largest among the four nearest tree individuals (*LargestF0*, *LargestF1*) were recorded in order to calculate CV, skewness and kurtosis.

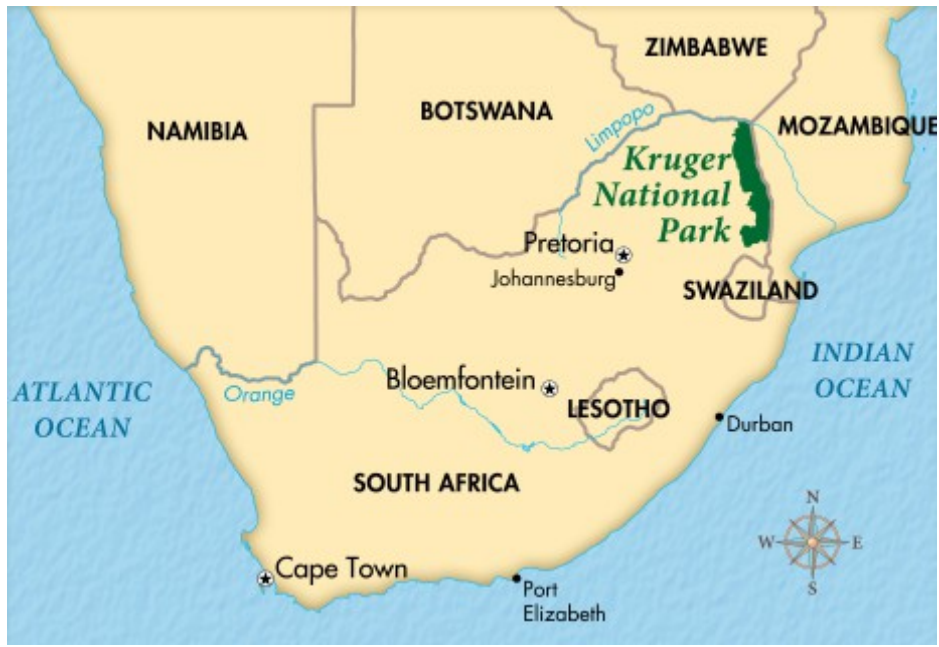

a

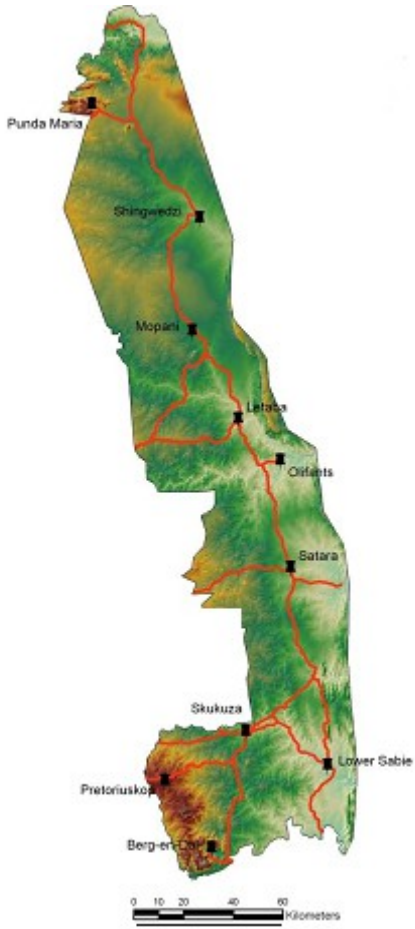

b

**Figure 1.** Geographic location of the study area. The study was conducted in the Kruger National Park (KNP) in South Africa. **a.** The location of the park in South Africa. **b.** Geophysical map of the park. The experimental plots sampled are all located within a maximum distance of 60 km from Skukuza.

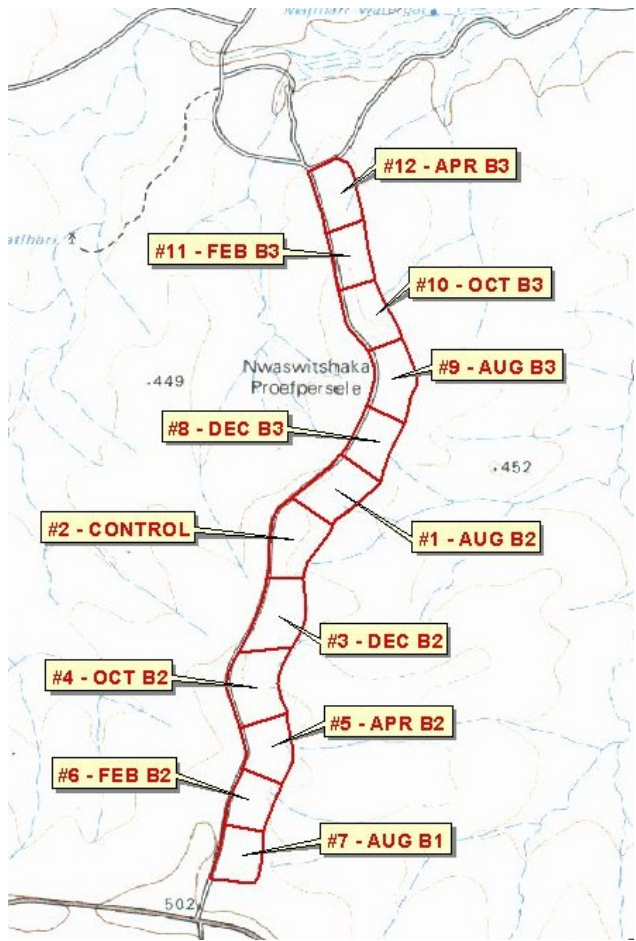

a.

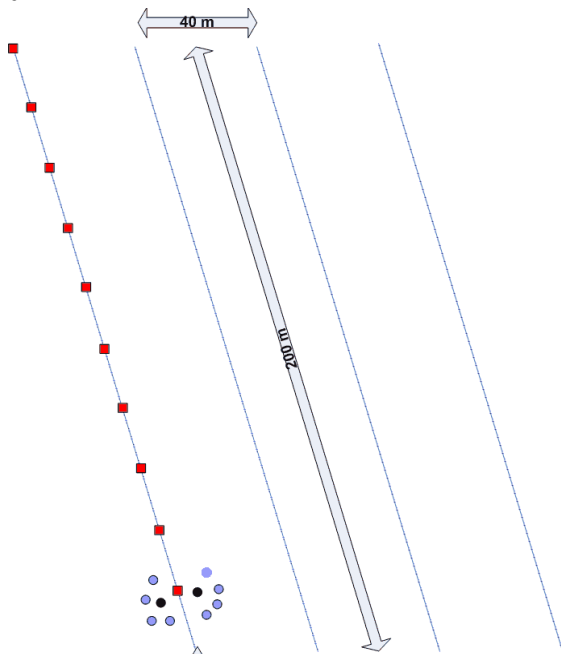

b

**Figure 2.** Overview of the experimental design. **a.** Overview of one (N'waswitshaka) out of four Experimental Burning Plots sampled. Each of the 12 plots shown in this figure has a surface area of 7 ha. There are 12 experimental burning treatments but only unburnt (*Control*) and annual burn every August (*Aug B1*) were sampled in this study. In the remaining 3 other EBPs sampled burning treatments are randomised in a split-plot design

and again only Control and Aug B1 burn were sampled. **b.** Overview of the sampling design of trees in each plot. Four line transects of 200 m each distancing 40 from each other were drawn. For each transect focal points (red squares) were drawn every 20 m on the line and the nearest two adult trees to the focal points were recorded (focal trees, black circles). The nearest four adult trees to each of the two focal trees were recorded (blue circles).

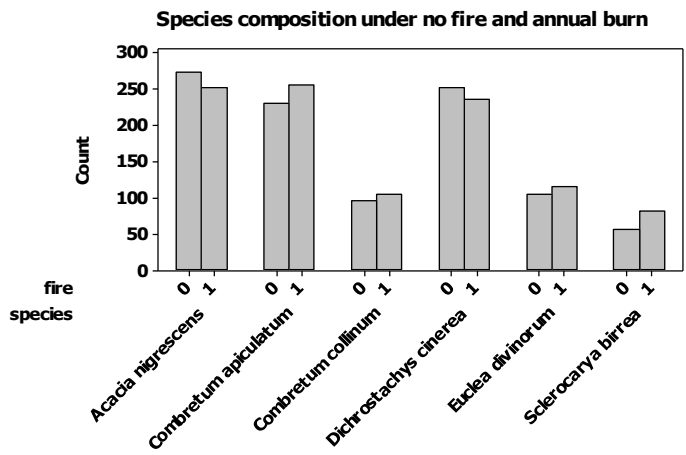

**Figure 3.** Counts of individuals of dominant tree species identities in unburnt (fire=0) and annual burn plots (fire = 1).

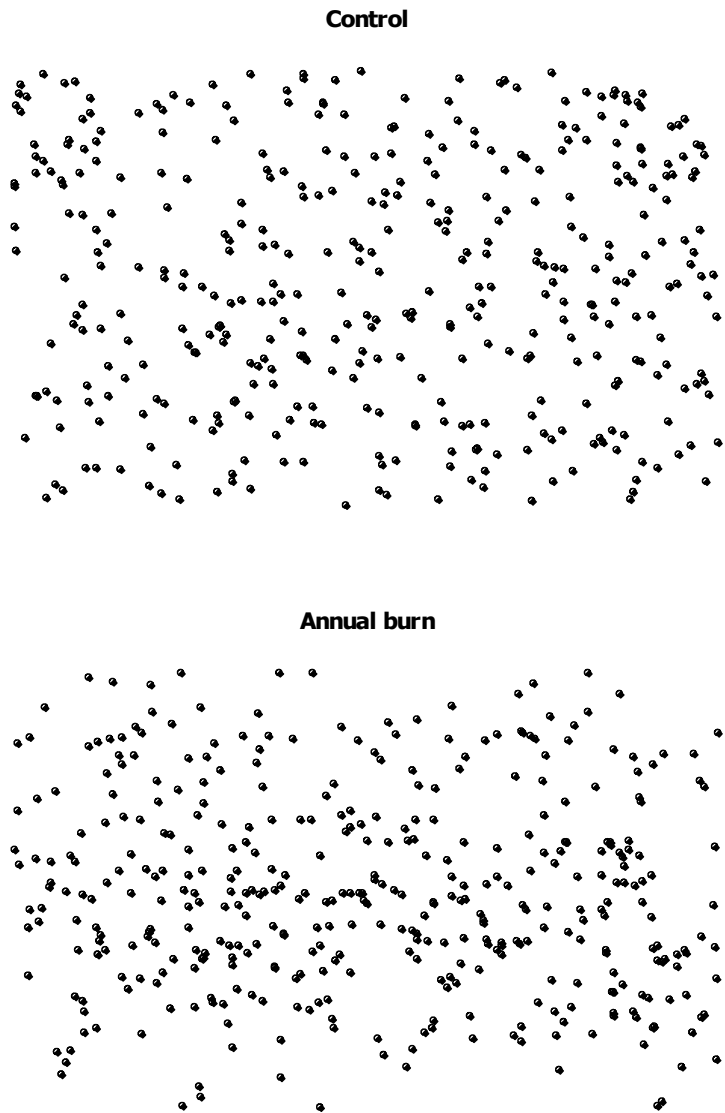

**Figure 4.** Spatial distribution of trees in one (N'waswitshaka) out of four Experimental Burning Plots sampled. Trees are plotted as points (their size in terms of TCAH is not plotted). **a.** Spatial distribution of trees in the control plot **b.** Spatial distribution of trees in the annual burn plot.

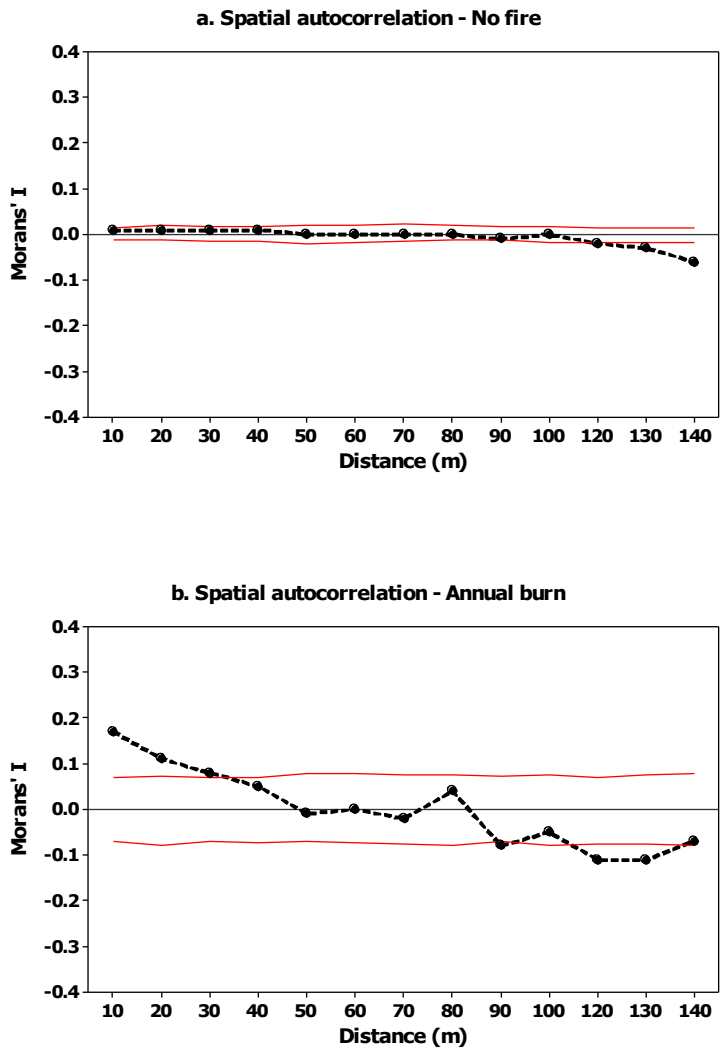

**Figure 5. (a)** Spatial autocorrelation values of tree distributions within unburned plots. **(b)** Spatial autocorrelation values of tree distributions within annual burn plots. Spatial autocorrelation was quantified in terms of values of Moran's I across distances (black dotted line). In order to calculate confidence intervals 999 randomisations were performed (upper and lower solid red lines). Values of Moran's I above the upper solid red line indicate significantly positive spatial autocorrelation, while values below the lower solid red line indicate significantly negative spatial autocorrelation. Values of Moran's I within the upper and lower solid red lines indicate spatial autocorrelation values that cannot be differentiated from random.

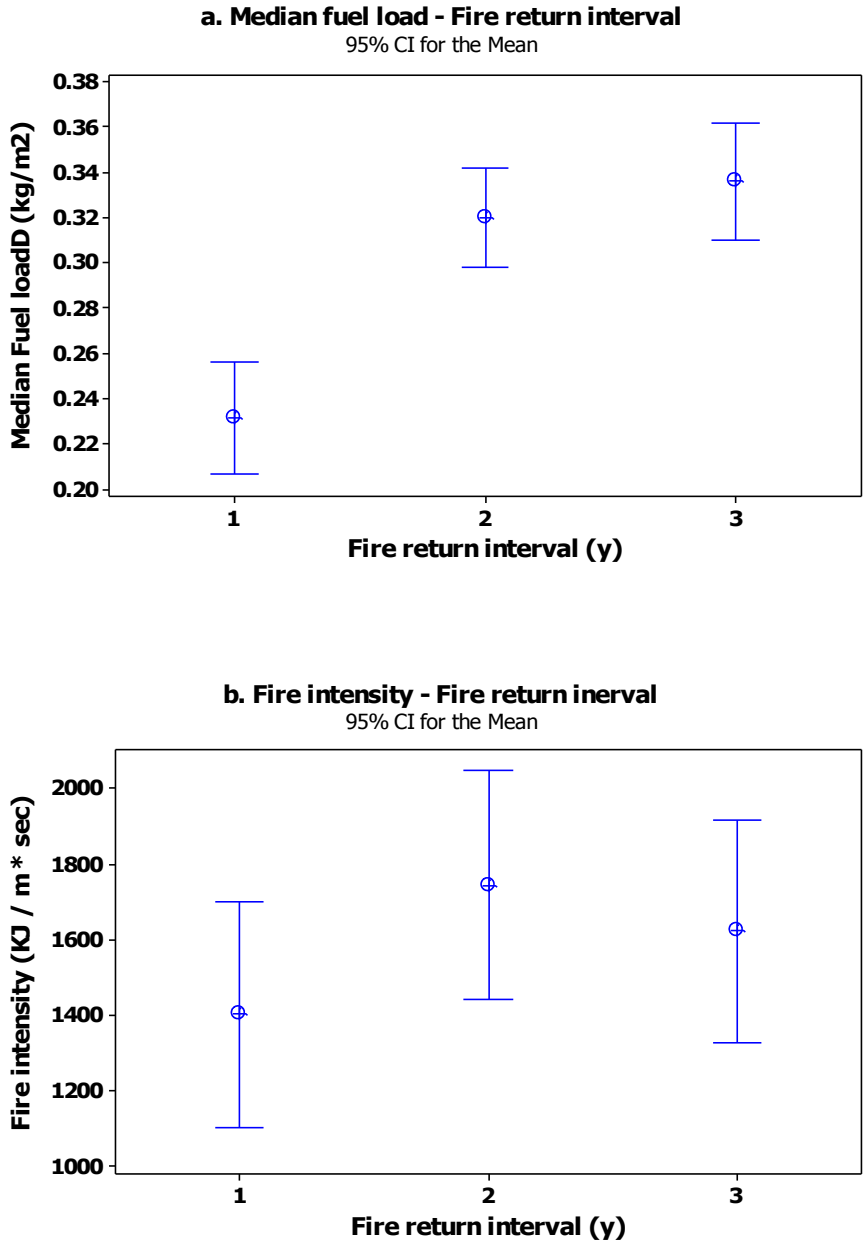

**Figure 6. (a)** Median fire fuel load in kilogram per square meter in experimental burning plots burned every year (y = 1), every two years (y = 2), and every three years(y = 3). **(b)** Fire intensity in kilo Joule per meter per second in experimental burning plots burned every year (y = 1), every two years (y = 2), and every three years(y = 3). Bars indicate 95% confidence intervals of the mean. Data from the Skukuza landscape in KNP, from with permission from Navashni Govender.